\documentclass{PoS}

\title{Lattice QCD approach to nuclear force}

\ShortTitle{Lattice QCD approach to nuclear force}

\author{\speaker{N.~Ishii}\\
        Center for Computational Sciences, University of Tsukuba,
	Tsukuba 305--8577, Japan\\
        E-mail: \email{ishii@ribf.riken.jp}}

\author{S.~Aoki\\
  RIKEN BNL Research Center, Brookhaven National Laboratory,
  Upton, New York 11973, USA\\
  Graduate School of Pure and Applied Sciences, University of Tsukuba,
  Tsukuba 305--8571, Japan\\
  E-mail: \email{saoki@het.ph.tsukuba.ac.jp}}

\author{T.~Hatsuda\\
  Department of Physics, University of Tokyo,
  Tokyo 113-0033, Japan\\
  E-mail: \email{hatsuda@phys.s.u-tokyo.ac.jp}}

\abstract{
We present our updated results of the nucleon-nucleon potential 
in  quenched lattice QCD simulations with the
plaquette  gauge action and  the Wilson  quark action  on the $32^4(\simeq
(4.4\mbox{fm})^4)$  lattice.
From the equal-time Bethe-Salpeter (BS)  wave function,
 the NN potential is constructed through the Schr\"odinger-type equation.
Resulting NN potential has all  the qualitative  features which
phenomenological potentials commonly  have: the repulsive  core at
short distance and the attractive well at medium and long distances.
In the $L\to\infty$  limit, our  NN potential is guaranteed to reproduce
the scattering length obtained from the L\"uscher's 
 formula.
The quark  mass dependence  of the NN  potential is  studied with
$m_{\pi}\sim 380,  529, 731$  MeV. The results  suggest that  both the
repulsive core at short distance and the attractive well  at medium distance
are enhanced in the light quark mass region.
}

\FullConference{The XXV International Symposium on Lattice Field Theory\\
		 July 30-4 August 2007\\
		 Regensburg, Germany}

\newcommand{\agt}{\raisebox{0.6ex}{$>$}\hspace*{-0.8em}\raisebox{-0.5ex}{$\sim$}}
\newcommand{\alt}{\raisebox{0.6ex}{$<$}\hspace*{-0.8em}\raisebox{-0.5ex}{$\sim$}}

\newcommand{\Eq}[1]{Eq.~(\ref{#1})}

\newcommand{\Fig}[1]{Fig.~\ref{#1}}

\begin{document}

\section{Introduction}
The  nuclear force  is the  essential ingredient  in  nuclear physics.
 Since H.~Yukawa  has introduced the  pion 72 yeas  ago \cite{yukawa},
 enormous efforts  have been devoted  to understand the origin  of the
 nuclear  force \cite{machleidt}.  At medium  to long  distances ($r
 \agt 1.2$  fm), the nuclear  force is attractive, which  is essential
 for the  existence of bound  nuclei.  Furthermore, the  nuclear force
 exhibits  the repulsive  core at  short distance  ($r \alt  0.7$ fm),
 which  is intimately  related to  the stability  of heavy  nuclei and
 neutron stars.  The  field theoretical treatment on the  basis of the
 pion and  heavy meson  exchanges has reasonable  success at  long and
 medium  distances.  However,  the repulsive  core at  short distance,
 which was first introduced by  R.~Jastrow 56 years ago to explain the
 high energy behavior  of the NN phase shifts  \cite{jastrow}, has not
 been understood with firm theoretical ground.
Since the nucleons begin to  overlap at short distances, the quark and
  gluon  structure of  the nucleon  is expected  to play  an important
  role.  In this respect,  the first principle lattice QCD calculation
  is desired to serve as the most powerful tool to attack the problems
  of the nuclear force.
There exits  a previous attempt to  apply the formalism  of the static
$q\bar{q}$ potential  to NN potential  \cite{takahashi}.  However, the
method  which  employs  the  static  quarks is  not  faithful  to  the
scattering data of nucleons.

 We  have  recently  generalized   the  method  developed  by  CP-PACS
 collaboration \cite{cp-pacs}  for the $\pi\pi$  scattering length and
 have carried out a first calculation of the NN potential from lattice
 QCD \cite{ishii}.   In our method, the  Bethe-Salpeter wave function,
 whose  asymptotic  form  leads  to  correct NN  phase  shift  in  the
 asymptotic  regime,  is  introduced  to construct  the  NN  potential
 through the Schr\"odinger-type equation.
The resulting NN potential has all the qualitative features
 required by phenomenology, i.e.,  the repulsive core at short distance
and the attractive well  at medium to long distances.
In the report,  we  extend our previous calculations 
by increasing  statistics and by introducing different quark masses 
corresponding to  the pion  masses,   $m_{\pi}
\simeq 380,  529, 731$ MeV. The  results show that  both the repulsive
core  and attractive well  are
enhanced in the light quark mass region.
%
%

\section{The formalism}
We begin  with the Schr\"odinger-type equation, which  is satisfied by
the Bethe-Salpeter wave function for the NN system,
\begin{eqnarray}
  \left(
  \vec{\nabla}^2 - k^2
  \right)
  \Psi_E(\vec r)
  &=&
  m_N
  \sum_{\vec r'} U(\vec r,\vec r') \Psi_E(\vec r')
  \label{schrodinger-like.eq}
  \\
  \Psi_{E,\alpha\beta}(\vec r)
  &\equiv&
  \lim_{t\to+0}
  \left\langle 0\left|
  T\left[p_{\alpha}(\vec{x},t) n_{\beta}(\vec y,0)\right]
  \right| NN(E) \right\rangle
  \label{bs.amplitude}
\end{eqnarray}
where  $\vec  r  \equiv  \vec   x  -  \vec  y$.   $p_{\alpha}(x)$  and
$n_{\beta}(y)$ denote  the standard local interpolating  fields for nucleons,
\begin{equation}
  p_{\alpha}(x)
  \equiv
  \epsilon_{abc}
  \left( u_a^T(x) C\gamma_5 d_b(x)\right) u_{c,\alpha}(x),
  \hspace*{2em}
  n_{\beta}(y)
  \equiv
  \epsilon_{abc}
  \left( u_a^T(y) C\gamma_5 d_b(y)\right) d_{c,\beta}(y),
  \label{standard.nucleon.op}
\end{equation}
where $a,b$  and $c$  denote the color  indices. $\alpha$  and $\beta$
denote the  Dirac indices. $C$ denotes the  charge conjugation matrix.
(For       derivation      of       \Eq{schrodinger-like.eq},      see
Refs.~\cite{cp-pacs,lin,aoki}.)  $U(\vec  r,\vec r')$, which  does not
depend  on $E$,  plays  a  role of  the  non-local interaction  kernel
\cite{aoki}.  The most general  (off-shell) form of NN potential after
imposing constraints  arising from  various symmetries is  analyzed in
Ref.~\cite{okubo}.   By  applying   the  derivative  expansion  up  to
$O(\vec\nabla^2)$, we obtain
\begin{eqnarray}
  U(\vec r,\vec r')
  &=&
  U_1(\vec r,\vec r')
  + (\vec\tau_1\cdot\vec\tau_2)
  U_{\tau\tau}(\vec r,\vec r')
  \\\nonumber
  &=&
  {\cal P}^{(I=0)} U^{(I=0)}(\vec r,\vec r')
  +
  {\cal P}^{(I=1)} U^{(I=1)}(\vec r,\vec r')
  \\\nonumber
  U^{(I)}(\vec r,\vec r')
  &=&
  V^{(I)} \cdot \delta(\vec x - \vec x')
  \\\nonumber
  V^{(I)}
  &=&
  V_0^{(I)}
  + (\vec\sigma_1\cdot\vec\sigma_2) V_{\sigma}^{(I)} 
  + S_{12} V_{\rm T}^{(I)}
  + (\vec L \cdot \vec S) V_{\rm LS}^{(I)}
  + \left\{\vec\sigma_1\cdot\vec L, \vec\sigma_2\cdot\vec L\right\}
  V_{\rm LL}^{(I)}
  + \left\{\vec\sigma_1\cdot\vec\nabla,\vec\sigma_2\cdot\vec\nabla \right\}
  V_{\rm pp}^{(I)}
  \\
  &\simeq&
  V^{(I)}_{\rm 0}(r)
  + (\vec \sigma_1\cdot\vec\sigma_2)V^{(I)}_{\sigma}(r)
  + S_{12} V^{(I)}_{\rm T}(r)
  + (\vec L\cdot \vec S) V^{(I)}_{\rm LS}(r)
  + O(\vec \nabla^2),
\end{eqnarray}
where ${\cal P}^{(I=0)}\equiv (1 - \vec\tau_1 \cdot \vec\tau_2)/4$ and
${\cal  P}^{(I=1)}\equiv (3 +  \vec\tau_1 \cdot  \vec\tau_2)/4$ denote
the projection matrices to $I=0$ and $I=1$ subspaces, respectively.
$\{*,*\}$ denotes the anti-commutator.
$S_{12}\equiv (\vec\sigma_1\cdot\vec r)(\vec\sigma_2\cdot\vec r)/r^2 -
\vec\sigma_1\cdot\vec\sigma_2$,     $\vec     L\equiv     -i     \vec
r\times\vec\nabla$  and   $\vec  S   \equiv  (\vec  \sigma_1   +  \vec
\sigma_2)/2$.
$V_{0}^{(I)}$,  $V_{\sigma}^{(I)}$, $V_{\rm  T}^{(I)}$, $V_{LL}^{(I)}$
and $V_{\rm  pp}^{(I)}$ are  functions of $\vec  r^2$, $\vec\nabla^2$,
and $\vec L^2$.
We combine the 1st and the 2nd terms in the last line as $V^{(I)}_{\rm
C}(r)\equiv   V^{(I)}_{0}(r)    +   (\vec   \sigma_1\cdot\vec\sigma_2)
V_{\sigma}^{(I)}(r)$, and  refer to  $V_{\rm C}^{(I)}(r)$ as  the {\em
central  force}.  $V_{\rm T}^{(I)}(r)$  and $V_{\rm  LS}^{(I)}(r)$ are
refereed  to  as  the {\em  tensor  force}  and  the {\em  LS  force},
respectively.   These three  forces play major roles in
conventional nuclear physics.

In QCD, the closest concept to the quantum mechanical wave function is
provided   by  the  equal-time    Bethe-Salpeter   (BS)   wave   function
\Eq{bs.amplitude}.    Note   that   \Eq{bs.amplitude}   represents   a
probability amplitude  to find  three quarks at  $\vec x$  and another
three   quarks  at   $\vec  y$.    It   is  possible   to  show   that
\Eq{bs.amplitude} has a proper asymptotic  behavior at $|\vec x - \vec
y| \to \infty$ \cite{aoki}. For example, in the $^1S_0$ channel, we have
\begin{equation}
  \Psi_{E}(\vec r)
  \rightarrow 
  e^{i\delta_0(k)}
  \frac{\sin\left(kr + \delta_0(k)\right)}{kr} .
  \label{bs.large.separation}
\end{equation}
The BS  wave function is  obtained from  the large  $t$ behavior  of the
``four-point'' correlator of the nucleon
\begin{eqnarray}
    \left\langle 0\left|
    T\left[
    p(\vec x,t)n(\vec x,t)
    W(t=t_0)
    \right]
    \right| 0 \right\rangle
  &=&
  \sum_{n}
  \left\langle 0\left|
  p(\vec x,0) n(\vec y,0)
  \right|n\right\rangle
  e^{-iE_n (t-t_0)}
  \left\langle n\left|
  W(t=0)
  \right|0\right\rangle.
\end{eqnarray}
Here,  $W(t)\equiv \bar P(t)  \bar N(t)$  represents the  wall source,
where  $P(t)$ and  $N(t)$ are  defined as  \Eq{standard.nucleon.op} with
 the quark  fields $\displaystyle  q(\vec  x,t)$ replaced by   $Q(t)\equiv
\sum_{\vec  x}q(\vec  x,t)$.   Since  contributions from  all  excited
states are  exponentially suppressed in  the large $t$ region,  we are
left with the BS wave function for the ground state.
In this report,  we  consider only $^1S_0$ and $^3S_1$ channels whose
   BS
wave functions  are 
\begin{eqnarray}
  \Psi(\vec r;^1S_0)
  &=&
  \frac1{24}
  \sum_{R\in O}
  \frac1{L^3}
  \sum_{\vec X}
  \left(\sigma_2\right)_{\alpha\beta}
  \left\langle 0 \left|
  p_{\alpha}(R\cdot\vec r + \vec X)
  n_{\beta}(\vec X)
  \right|NN(E)\right\rangle 
  \\\nonumber
  \Psi(\vec r;^3S_1)
  &=&
  \frac1{24}
  \sum_{R\in O}
  \frac1{L^3}
  \sum_{\vec X}
  \left(\sigma_2 \sigma_3\right)_{\alpha\beta}
  \left\langle 0 \left|
  p_{\alpha}(R\cdot\vec r + \vec X)
  n_{\beta}(\vec X)
  \right|NN(E)\right\rangle,
\end{eqnarray}
where  the   summations  over  $R\in   O$  are  performed   for  cubic
transformation group.   The summations for  $\vec X$ are  performed to
select the zero total spatial momenta.

First we  consider the Schr\"odinger equation in $^1S_0$ channel. Owing
to the identical  nature of the nucleons,  two nucleon system
in $^1S_0$ channel is  iso-vector. Since the contributions
from the tensor  force and the LS force vanish in this channel,  we  are left with the
following Schr\"odinger equation,
\begin{equation}
  \left(
  -
  \frac1{2\mu}
  \vec\nabla^2
  +
  V^{(I=1)}_{\rm C}(r)
  \right)
  \Psi(\vec r; ^1S_0)
  =
  \frac{k^2}{2\mu}
  \Psi(\vec r; ^1S_0),
\end{equation}
where $\mu\equiv m_N/2$ denotes the  reduced mass of the nucleon.  $k$
plays  the   role  of  the  ``{\em   asymptotic  momentum}''.   Since
$V^{(I=1)}_{\rm  C}(r)=V^{(I=1)}_{0}(r) - 3  V^{(I=1)}_{\sigma}(r)$ is
an ordinary function,  which does not involve a  derivative nor matrix
structure, we arrange the Schr\"odinger equation to obtain
\begin{equation}
  V^{(I=1)}_{\rm C}(r)
  =
  \frac{k^2}{2\mu}
  +
  \frac1{2\mu}
  \frac{\vec\nabla^2 \Psi(\vec r; ^1S_0)}{\Psi(\vec r; ^1S_0)}.
  \label{cp-pacs.koshiki}
\end{equation}

Next, we consider the  Schr\"odinger equation in $^3S_1$ channel. 
 In this case,   two nucleon system
 is iso-scalar.  Unlike $^1S_0$  case, this channel
receives  a non-vanishing  contribution from  the tensor  force, which
provides a  coupling to $^3D_1$  channel.  $^3D_1$ channel  receives a
contribution from the LS force.   In this way, we have three unknowns,
i.e.,    $V^{(I=0)}_{\rm    C}(r)$,    $V^{(I=0)}_{\rm   T}(r)$    and
$V^{(I=0)}_{\rm LS}(r)$ with two equations, i.e., $^3S_1$ and $^3D_1$.
($V^{(I=0)}_{\rm   C}(r)=V^{(I=0)}_{0}(r)   +   V^{(I=0)}_{\sigma}(r)$
should not be confused with $V^{(I=1)}_{\rm C}(r)$.)
To obtain these  three forces exactly, we need  one more equation
 such as the Schr\"odinger equation in $^3D_2$ channel.
 In this report, we do not  pursue this direction.  Instead, we
adopt the same procedure as  the $^1S_0$ channel. This leads to
 the  so-called  ``{\em
effective central  force}'' $V_{\rm C}^{\rm eff}(r)$  which  takes into
account the  $^3D_1$ channel  indirectly through the 
tensor force.

\section{The lattice QCD result}

We  employ  the  standard  plaquette  action on  $32^4$  lattice  with
$\beta=5.7$ to generate quenched gauge configurations.
The gauge configurations are picked up every 200 sweeps after skipping
3000 sweeps for thermalization.
Quark propagators are generated by employing the standard Wilson quark
action with $\kappa=0.1640, 0.1665$ and $0.1678$.
%
The scale  unit $1/a=1.44(2)$ GeV  ($a\simeq 0.137$ fm)  is introduced
from the rho meson mass in the chiral limit \cite{fukugida}.
The physical size of our lattice corresponds to $L\sim 4.4$ fm.
The  number of  gauge configurations  $N_{\rm conf}$,  the  pion mass
$m_{\pi}$,   the  nucleon   mass   $m_{\rm  N}$   are  summarized   in
Table~\ref{table}.  (For $\kappa=0.1678$,  24 gauge configurations are
identified as  exceptional configurations, which  are not used  in the
calculations.)
\begin{table}
\begin{center}
\begin{tabular}{cccccll}
\hline
$\kappa$ & $N_{\rm conf}$ & $m_{\pi}$ [MeV] & $m_{\rm N}$ [MeV] & $t-t_0$
& $E(^1S_0)$ [MeV] & $E(^3S_1)$ [MeV] \\
\hline
0.1640   & 1000           & 732.1(4) & 1558.4(63) & 7 & $-0.400(83)$ & $-0.480(97)$\\
0.1665   & 2000           & 529.0(4) & 1333.8(82) & 6 & $-0.509(94)$ & $-0.560(114)$\\
0.1678   & 2021           & 379.7(9) & 1196.6(32)& 5  & $-0.675(264)$ & $-0.968(374)$\\
\hline
\end{tabular}
\end{center}
\caption{The number  of gauge configurations $N_{\rm  conf}$, the pion
mass $m_{\pi}$,  the nucleon mass  $m_{\rm N}$, time-slice  $t-t_0$ on
which  BS  wave  functions  are  measured,  and  the  non-relativistic
energies $E\equiv k^2/(2\mu)$ for $^1S_0$ and $^3S_1$ channels.}
\label{table}
\end{table}

The  periodic (Dirichlet) boundary  condition is  imposed on  the quark
fields along the spatial (temporal) direction. We adopt the wall source
on the time-slice $t=t_0\equiv 5$.  The BS wave functions are measured
on  the time-slice  $t-t_0  =  7, 6,  5$  for $\kappa=0.1640,  0.1646,
0.1678$,  respectively.  The  ground state  saturation is  examined by
t-dependence  of  the NN  potential.   We  employ the  nearest
neighbor representation of  the discretized Laplacian as $\vec\nabla^2
f(\vec x) \equiv \sum_{i=1}^{3}\left\{ f(\vec  x + a\vec e_i) + f(\vec
x -  a\vec e_i)\right\} - 6  f(\vec x)$, where $\vec  e_i$ denotes the
unit vector  along the $i$-th  coordinate axis. BS wave  functions are
fully measured for  $|\vec r| \alt 0.7$ fm, where  rapid changes of BS
wave function and  NN potential are expected. 
Since  the  changes  are  rather  modest for  $|\vec r| \agt 0.7$
fm,   the
measurement of BS wave functions  is restricted on the coordinate axes
and their  nearest neighbors to  reduce the calculational  cost. 
%
The ``asymptotic momentum'' $k^2$ is  obtained by fitting the BS wave
function:  We use the Green's function of the Helmholtz equation,
\begin{equation}
  G(\vec   r;k^2)
  \equiv
  \frac1{L^3}
  \sum_{\vec   n\in\mathbb{Z}^3}
  \frac{\exp\left(i 2\pi  \vec n\cdot\vec r/L\right)}{(2\pi/L)^2\vec  n^2 - k^2},
\end{equation}
as  the fit  function by  regarding the  overall numerical  factor and
$k^2$ as fit parameters.  (An appropriate regularization is assumed in
this representation of Green's function.)
The fits are  performed outside of the range  of NN interaction, which
is  determined by  examining  $\vec\nabla^2\Psi(\vec x)/\Psi(\vec  x)$
\cite{cp-pacs}.

\begin{figure}
\begin{center}
\includegraphics[width=0.34\textwidth,angle=-90]{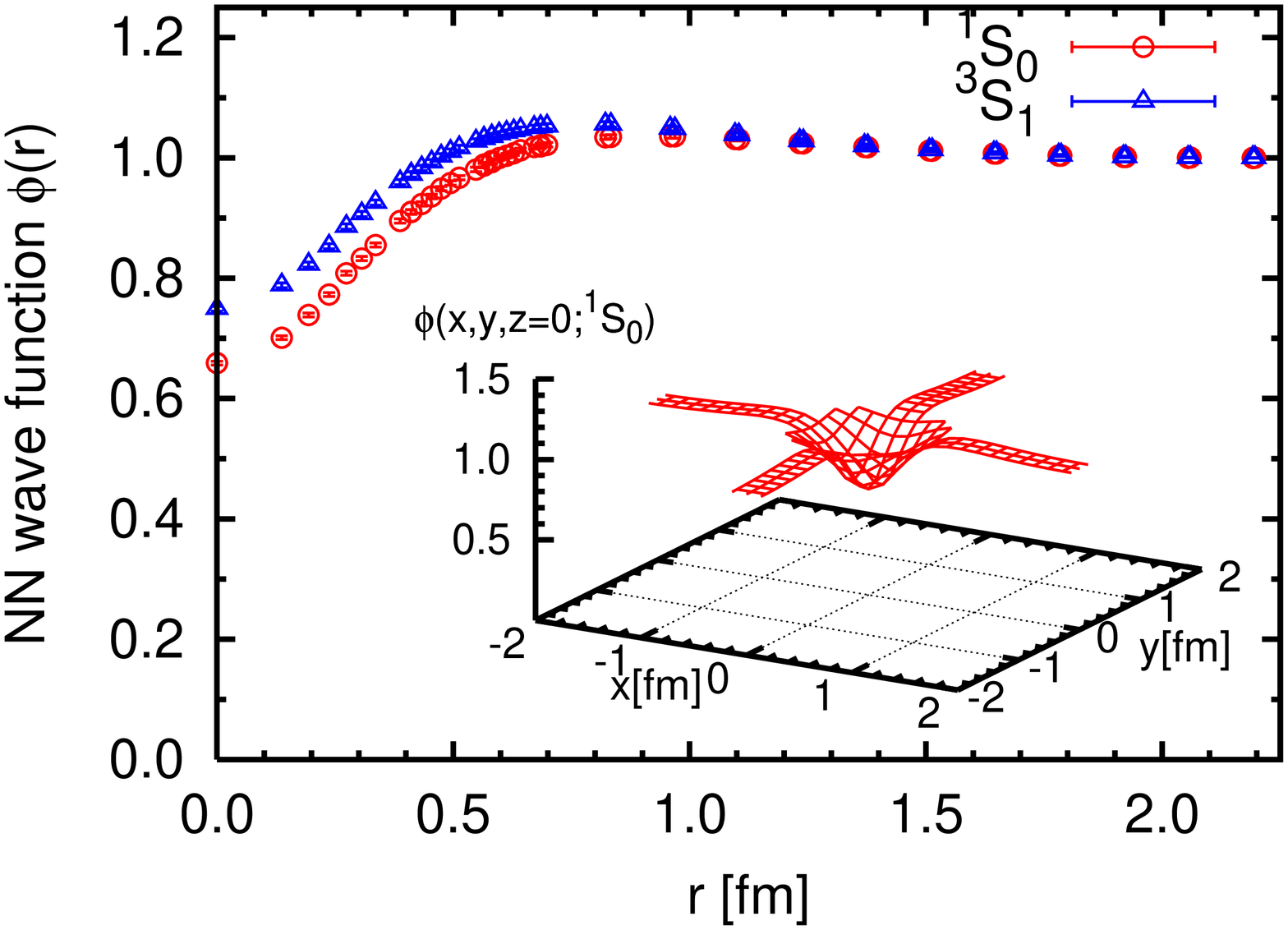}
\includegraphics[width=0.34\textwidth,angle=-90]{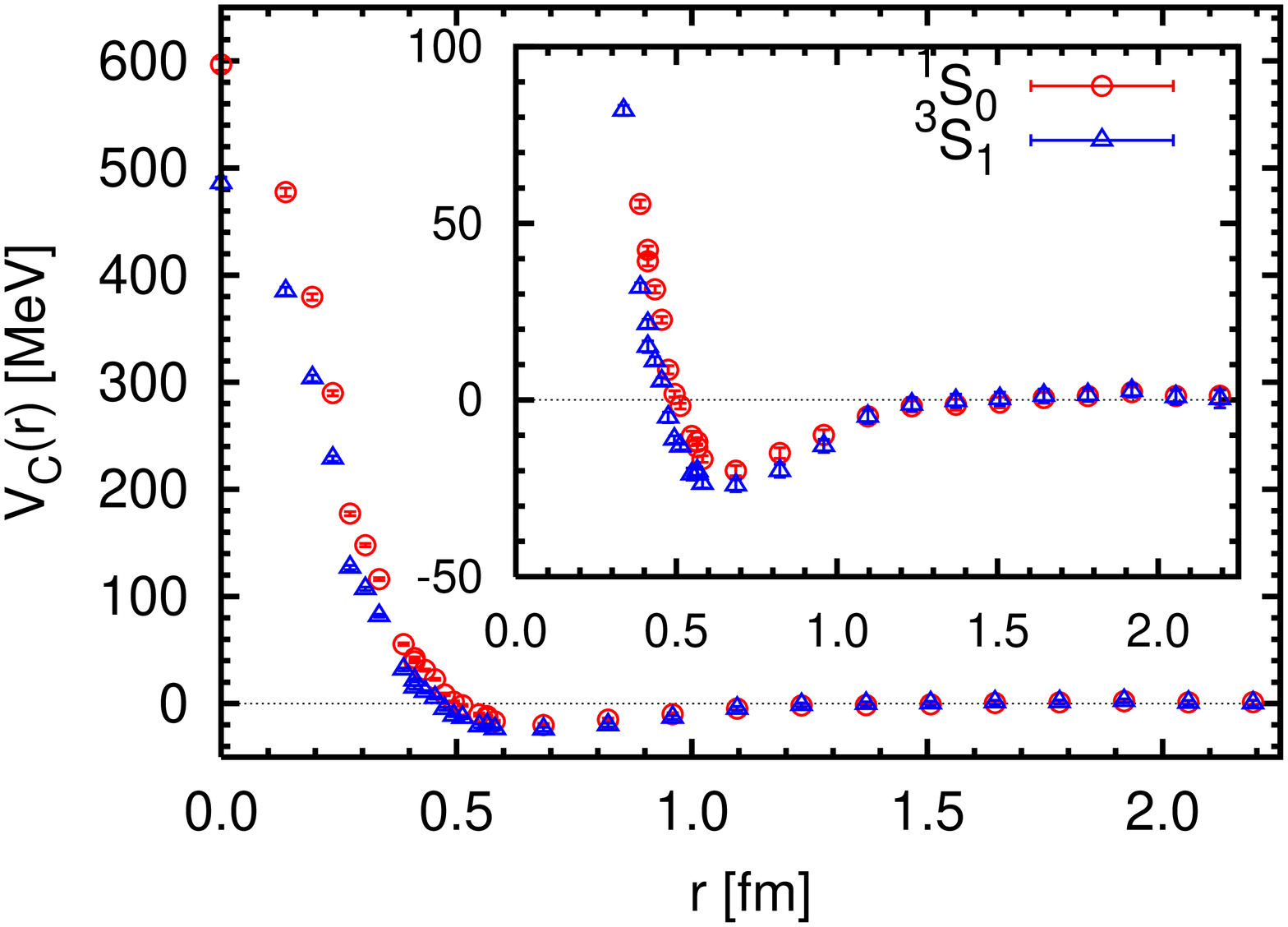}
\end{center}
\caption{NN wave functions in $^1S_0$ and $^3S_1$ channels (left), and
NN  potentials (right),  i.e., central  force in  $^1S_0$  channel and
effective central force in $^3S_1$ channel for $\kappa=0.1665$.
The  inset of  the  left figure  is a  3D  plot of  the wave  function
$\phi(x,y,z=0; ^1S_0)$.}
\label{fig1}
\end{figure}

\Fig{fig1} (left) shows  the BS wave functions in  $^1S_0$ and $^3S_1$
channels for $\kappa=0.1665$.  The suppression of the wave function in
the region $r\alt 0.5$ fm indicate the existence of repulsion at short
distance.   \Fig{fig1} (right) shows  the reconstructed  NN potentials
for $\kappa=0.1665$,  i.e., the central force for  $^1S_0$ channel and
the   effective    central   force   for    $^3S_1$   channel.    (See
Table~\ref{table},  for the  values of  the  non-relativistic energies
$E\equiv k^2/(2\mu)$ in \Eq{cp-pacs.koshiki}.)
We see that our NN potentials have repulsive cores of $500-600$ MeV in
the short distance region ($r\alt 0.5$ fm) and attractions of about $30$
MeV in the medium distance region ($0.5 \alt r \alt 1.2$ fm).
Both the repulsive core at short distance and the attractive well at medium
distance are weaker than those expected phenomenologically.
This is due to the heavy  quark mass in our simulations  In the  light quark mass
region, pion can propagate longer distance, which is expected to enhance
the attraction  at medium and  long distance.  

\begin{figure}
\begin{center}
\includegraphics[width=0.60\textwidth,angle=-90]{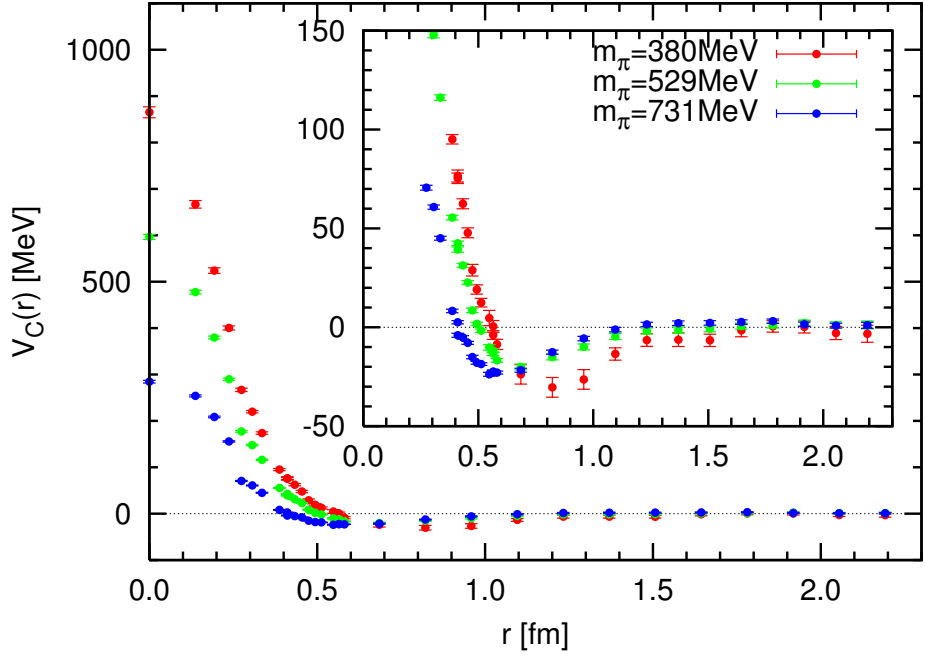}
\end{center}
\caption{
Central forces in $^1S_0$ channel for three quark masses.
}
\label{fig2}
\end{figure}
 To see  the quark mass dependence  in a quantitative  manner, we show
 the  NN  potentials  for  three  different quark  masses  in  $^1S_0$
 channel in \Fig{fig2}.
 As the quark  mass decreases, the repulsive core at short
distance  is  enhanced  rapidly,  whereas  the  attraction  at  medium
distance is  modestly enhanced.   This indicates that  it is  important to
perform the lattice QCD calculation  in the lighter quark mass region in
order to compare our result with experimental data.

Several comments are in order here.\\
(i)  The net interaction  is attractive  even in  the presence  of the
repulsive core.  Indeed, L\"uscher's finite volume method leads to the
attractive    scattering    length,   i.e.,    $a_0(^1S_0)=0.115(26)$,
$0.126(25)$,  $0.159(66)$ fm and  $a_0(^3S_1)=0.140(31)$, $0.140(31)$,
$0.252(104)$  fm for  $\kappa=0.1640,  0.1665, 0.1678$,  respectively.
The attractive nature of our potential is 
qualitatively  understood by the Born  approximation formula for  the scattering
length
$
  a_0
  \simeq
  -m_N\int V_{\rm C}(r) r^2 dr.
$
Owing to the volume factor $r^2 dr$, the attraction at medium distance
overcomes the repulsive core at short distance.
\\
(ii)  There is  a considerable discrepancy  between
the  above scattering   lengths   and   the  empirical   values,   i.e.,
$a_0^{\rm(exp)}(^1S_0)$ $\sim$  $20$ fm and $a_0^{\rm(exp)}(^3S_1)\sim
-5$ fm.   This is attributed to the heavy quark mass employed in our simulations.
 If we  can get closer  to the physical  quark mass, there  appears an
 ``unitary region" where the  NN scattering length becomes singular as
 a    function    of    the    quark    mass    and    changes    sign
 \cite{kuramashi,beane}.  The   singular  point  is   related  to  the
 threshold  of  bound  state  formation.   This is  why  the  physical
 scattering  length is  positively large  in the  $^1S_0$  channel (no
 bound state) and is negatively large in the $^3S_1$ channel (deuteron
 bound state) .

\section{Summary}

We  have extended our previous results of the nuclear force on the 
lattice by increasing statistics and adopting different quark masses.
The NN  potentials in the $^1S_0$ and $^3S_1$ channels
have   all  the  qualitative   features  which   phenomenological  NN
potentials commonly have, i.e.,  the repulsive core 
 at short distance and attractive well at medium to long distances.
The quark mass dependence of the NN potential shows that
 the repulsive  core  at  short  distance  is  enhanced  rapidly,  and  the
attraction at medium distance is modestly enhanced.
These results suggest  that, in order to compare  our results with the
experimental data,  it is important  to perform the lattice  QCD Monte
Carlo calculation in the lighter quark mass region.
 
  Although  the BS  wave functions  are proved  to have  the universal
behavior  as  \Eq{bs.large.separation} at  large  distance, its  short
distance behavior  is afflicted with  the operator dependence.  We can
avoid  this subtlety by  resorting to  the inverse  scattering theory,
which  guarantees the unique  existence of  energy-independent, local,
hermitian  potential.  Studies along  this line  together with  the NN
potential measured in  different $E$ and the derivation  of the tensor
force will be reported elsewhere \cite{aoki}.

\section*{Acknowledgements}
Lattice QCD Monte Carlo calculation has been done with IBM Blue Gene/L
at  KEK.   This research  was  partly  supported  by the  Ministry  of
Education, Science,  Sports and Culture,  Grant-in-Aid (Nos. 13135204,
15540251, 15540254, 18540253, 19540261).

\end{document}